\documentclass[aps,prl,twocolumn,superscriptaddress]{revtex4}
\usepackage{graphicx}
\usepackage{setspace}
\usepackage{amsmath}
\usepackage{dcolumn}
\usepackage{color}

\newcommand{\bom}{\mbox{\boldmath $\bf\omega$}}

\def\V{\textbf{V}}
\def\bv{\textbf{v}}
\def\B{\textbf{B}}
\def\bh{\textbf{h}}
\def\e{\textbf{e}}
\def\Ha{\textrm{Ha}}
\def\Re{\textrm{Re}}
\def\Pr{\textrm{Pr}}
\def\R{\textrm{R}}
\def\pa{\partial}

\begin{document}

\title{Magnetorotational instability in electrically driven fluids}

\author{I. V. Khalzov}
\affiliation{University of Saskatchewan, Saskatoon, Canada}
\affiliation{Russian Research Center "Kurchatov Institute", Moscow,
Russia.}
\author{A. I. Smolyakov}
\affiliation{University of Saskatchewan, Saskatoon, Canada}
\affiliation{Russian Research Center "Kurchatov Institute", Moscow,
Russia.}
\author{V. I. Ilgisonis}
\affiliation{Russian Research Center "Kurchatov Institute", Moscow,
Russia.}

\date{\today}

\begin{abstract}
The linear stability of electrically driven flow of liquid metal in
circular channel in the presence of vertical magnetic field is
studied. It is shown that the instability threshold of such flow is
determined by magnetorotational instability of non-axisymmetric
modes ($m\neq0$) and does not depend on the type of the fluid if
magnetic Prandtl number is small $\Pr\ll1$. Our numerical results
are found to be in a good agreement with available experimental data
from Grenoble High Magnetic Field Laboratory, France [P. Moresco and
T. Alboussi\`{e}re, J. Fluid Mech. \textbf{504}, 167 (2004)].
\end{abstract}

\maketitle

The theoretical and experimental study of  magnetorotational
instability (MRI) has attracted much attention in recent years. MRI
was originally predicted by Velikhov \cite{Vel} in 1959, but its
intensive study began only in 1991  when it was rediscovered in
astrophysical context by Balbus and Hawley \cite{BH}.
At present time subject of MRI is one of the most important
developments in magnetohydrodynamic (MHD) theory with far reaching
consequences for a variety of astrophysical phenomena such as
accretion disks \cite{BH_rev,Balbus_rev}, magnetic reconnection
\cite{Coppi} and dynamo \cite{Hughes}.

One of the topics of great current interest is experimental
verification of MRI. Several experiments have been initiated to
investigate MRI in laboratory \cite{PPPL, Los1, Rud, Maryland, RRC,
Los2} by studying the stability of conducting fluid (liquid metal)
rotating in transverse magnetic field. Despite these attempts, the
MRI has never been clearly detected in laboratory and any progress
in this direction is extremely important. Current status of
experimental studies of MRI is described in \cite{Balbus2,Good}.


Two different mechanisms of the fluid rotation have been proposed
for MRI experiments so far: mechanical drive by virtue of viscous
drag force acting on the fluid between moving surfaces (Couette
flow) and electrical drive by Ampere force arising when the electric
current is passed through the fluid in transverse magnetic field
(electrically driven flow). In most existing MRI experiments a
Couette flow is used either  in cylindrical \cite{PPPL, Los1, Rud}
or spherical geometry \cite{Maryland}. The main difficulty of the
cylindrical Couette flow is the presence of the stationary end-caps
that affect the entire equilibrium flow making it different from the
idealized infinite-cylinder angular velocity profile
$\Omega(r)=a+b/r^2$ so the conditions for MRI may not be met. Their
influence can be reduced either by employing  the differentially
rotating end-caps \cite{Good,PPPL2} or by using sufficiently long
cylinders. In the latter case, experimental observations  of MRI
have been reported \cite{Rud, Donn}. MRI has also been observed in
the experiment with rotating spheres \cite{Maryland} though the
background flow was already fully turbulent without any magnetic
field, indicating that MRI is not the only possible instability in
this geometry.

\begin{figure}[tb]
  \includegraphics[scale=0.35]{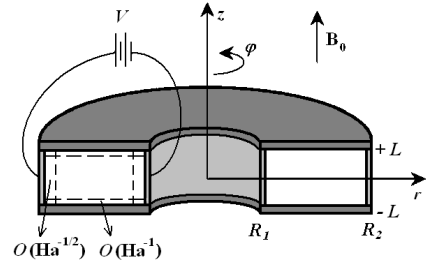}\\
  \caption{Sketch of electrically driven flow.
  Some details are shown as in Fig. 1 in Ref. \cite{Grenoble}.}\label{fig1}
\end{figure}

Another way to rotate conducting fluid in circular channel with
axial magnetic field is  to  apply  radial electric current
(Fig.~\ref{fig1}). In this case the equilibrium flow forms so-called
Hartmann layers near the top and bottom walls and parallel boundary
layers near the side walls. The widths of these layers scale with
Hartmann number $\Ha$ (see Eq. (\ref{param}) for definition) as
$O(\Ha^{-1})$ and $O(\Ha^{-1/2})$ respectively, so they become
negligible at high values of magnetic field \cite{Hunt}. Such flow
has the angular velocity profile $\Omega(r)\propto 1/r^2$ almost
entirely in the cross-section of the channel \cite{my1}. This
profile is stable in hydrodynamics according to Raleigh's criterion,
but it can be destabilized in the presence of magnetic field since
it satisfies the necessary condition for MRI $\pa\Omega^2(r)/\pa
r<0$ \cite{Vel}. MRI experiment based on electrically driven flow of
liquid sodium has been proposed in the Russian Research Center
"Kurchatov Institute" \cite{RRC} and built in Obninsk (Russia). A
similar configuration using plasma instead of liquid metal has been
developed in Los Alamos \cite{Los2}. At present time experimental
data are not available from those MRI experiments.

The  electrically driven flow of liquid metal (mercury) in circular
channel was also used in Grenoble High Magnetic Field Laboratory,
France to investigate the stability of the Hartmann layers
\cite{Grenoble}. In this experiment a well-marked transition to
turbulence was found when the ratio of Reynolds number to Hartmann
number $\R=\Re/\Ha$ exceeded critical value $\R_c^*\approx380$ (star
denotes the quantities taken from \cite{Grenoble}), this value of
$\R_c^*$ was valid also for inverse process of laminarization
without a visible hysteresis and for a wide range of intensities of
the magnetic field ($\Ha=100\div1700$). The critical value $\R_c^*$
obtained in Grenoble experiment is two orders of magnitude smaller
than $\R_c^{\Ha}\approx50,000$ predicted by the linear stability
theory for Hartmann layers \cite{Lock}. A possible explanation to
this paradox is discussed in papers \cite{Kras1,Kras2}, where a
complicated two-step transition scenario to turbulence in the
Hartmann layers is assumed. In these papers the radial dependence of
the velocity profile in the main part of the flow is completely
disregarded. In the present Letter, it is suggested  however that in
this configuration the magnetized flow can be destabilized via the
MRI mechanism.

We study the linear stability of electrically driven flow of liquid
metal in the geometries relevant to the Grenoble and Obninsk MRI
experiments (Table~\ref{table1}) and show that the instability
thresholds in these experiments are determined by MRI of global
non-axisymmetric modes with large azimuthal mode numbers $m$. In our
analysis we neglect all boundary layers and assume that the fluid
rotates in the  uniform axial magnetic field $\B_0=B_0\e_z$ with
equilibrium velocity $\V_0=r\Omega(r)\e_{\varphi}$ where
$\Omega(r)=M_0/r^2$ throughout the cross-section of the channel. The
equilibrium angular momentum of the velocity $M_0$ is determined by
the total electric current $I_0$ passing through the channel
\cite{my1}:
$$
M_0=\frac{I_0}{4\pi\sqrt{\rho\sigma\nu}},
$$
where $\rho$, $\sigma$ and $\nu$ are fluid density, electric
conductivity and kinematic viscosity respectively.

\begin{table}[tb]
\caption{Parameters of Grenoble and Obninsk experiments.
\label{table1}}
\begin{ruledtabular}
\begin{tabular}{cccccc}
& $R_1$ (cm) & $R_2$ (cm) & $L$ (cm) & Fluid & $\Pr$ \\
\hline
Grenoble & 4 & 5 & 0.5 & Hg & $1.49\cdot10^{-7}$\\
Obninsk  & 3 & 15 & 3 & Na & $8.8\cdot10^{-6}$ \\
\end{tabular}
\end{ruledtabular}
\end{table}

A number of similar studies was performed by different researches
\cite{PPPL,Los1,Rud2,Rud3}, who considered the stability of more
general idealized Couette flow $\Omega(r)=a+b/r^2$. The major part
of these studies is restricted to the case of axisymmetric modes
with $m=0$ \cite{PPPL,Los1}, which are known to have the largest
growth rate of MRI. Numerical analysis of non-axisymmetric modes
with small $m$ \cite{Rud2,Rud3} shows that they can have lower
instability threshold (lower critical values of Reynolds number
$\Re$), so they might be easier to excite in real experiment. For
electrically driven flow it was found recently \cite{my2} that in
ideal MHD the MRI threshold decreases with $m$ as $\Re\propto1/m$;
therefore the most dangerous modes in such flow are those with
larger $m$ (though they have smaller instability growth rate). Our
present results suggest that overall stability of the flow in a
finite height circular channel is determined by dissipative
perturbations with large azimuthal numbers $m$.

We use dissipative incompressible MHD equations linearized about the
equilibrium state and represent all perturbations in the form
\hbox{$f(r,z)\exp(\gamma t + im\varphi)$} in cylindrical system of
coordinates $\{r,\varphi,z\}$. It should be stressed here that we
consider the channel of the finite height -- this is a substantial
difference of our stability analysis from previous studies. For
computational convenience we introduce dimensionless quantities,
taking as a unit of length a half-height of the channel $L$ and as a
unit of time the characteristic viscous time $L^2/\nu$. The
perturbations of velocity $\delta\V$ and magnetic field $\delta\B$
can be written in terms of dimensionless vectors $\bv$ and $\bh$ as
$$
\delta\V=\frac{M_0}{L}\bv,~~~\delta\B=B_0\frac{\Pr\Re}{\Ha}\bh,
$$
where
\begin{equation}
\label{param} \Ha=\frac{LB_0}{c}\sqrt{\frac{\sigma}{\rho\nu}},~~~
\Re=\frac{M_0}{\nu},~~~ \Pr=\frac{4\pi\sigma\nu}{c^2}
\end{equation}
are Hartmann, Reynolds and magnetic Prandtl numbers respectively.
Introducing vortex $\bom=\nabla\times\textbf{v}$ we arrive at
\begin{eqnarray}
\label{main1} \gamma{\bom}&=&\nabla^2\bom  - \frac{im\Re}{r^2}\bom -
\frac{2\Re}{r^2}\omega_r\e_{\varphi} + \Ha\nabla\times\textbf{h}'_z,\\
\label{main2} \Pr\gamma\textbf{h}&=&\nabla^2\textbf{h} -
\frac{im\Pr\Re}{r^2}\textbf{h} - \frac{2\Pr\Re}{r^2}h_r\e_{\varphi}
+  \Ha\textbf{v}'_z,
\end{eqnarray}
where prime denotes the derivative with respect to $z$. Note that
$r$ and $\varphi$ components of these equations are enough to find
the full solution ($z$-components can be deduced from the conditions
$\nabla\cdot\bom=0$ and $\nabla\cdot\bh=0$).

The proper boundary conditions should be specified prior to solving
the system (\ref{main1}), (\ref{main2}). In the flow of viscous
fluid all velocity components vanish at the rigid walls, i.~e.
\begin{equation}
\label{bcV} \textbf{v}|_{r=r_1,r_2}=0,~~~\textbf{v}|_{z=\pm1}=0,
\end{equation}
The boundary conditions for magnetic field depend on the
conductivity of the walls. In both Grenoble and Obninsk experiments
the side walls of the channel can be considered as perfect
conducting. At the surface of the perfect conductor the time-varying
normal component of magnetic field as well as the tangential
components of electric current should be zero. This means
\begin{equation}
\label{bcHr} h_r|_{r=r_1,r_2}=0,~~~\frac{\pa (rh_{\varphi})}{\pa
r}\bigg|_{r=r_1,r_2}=0.
\end{equation}
The other two walls (Hartmann walls) are electrical insulators.
For simplicity we assume that the perturbed components of the field
are zero at these walls, i. e.
\begin{equation}
\label{bcHz} h_r|_{z=\pm1}=0,~~~h_{\varphi}|_{z=\pm1}=0,
\end{equation}
which is consistent with the absence of the normal component of the
current at the surface of insulator.

Equations (\ref{main1}), (\ref{main2})  with boundary conditions
(\ref{bcV})-(\ref{bcHz}) constitute an eigenvalue problem with
$\gamma$ being an unknown eigenvalue. A solution to this problem was
sought by expanding functions $v_r$, $v_{\varphi}$, $h_r$ and
$h_{\varphi}$ in terms of either odd or even polynomials in $z$ (up
to $N_z=8$), and by discretization of the system (\ref{main1}),
(\ref{main2}) in terms of finite differences in $r$-direction (up to
$N_r=50$). Then this system was reduced to a large
($4N_rN_z\times4N_rN_z$) matrix eigenvalue problem which was solved
using standard numerical methods of MATLAB.

\begin{figure}[tb]
  \includegraphics[scale=0.35]{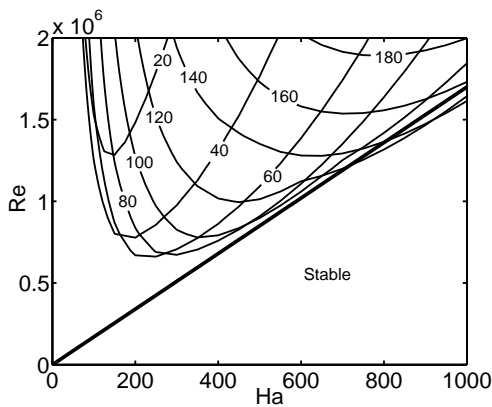}\\
  \caption{Calculated
  marginal stability curves for geometry of Grenoble experiment.
  Azimuthal mode numbers $m$ are shown. Difference between Hg and Na is within the line width.
  Straight line corresponds to the
   transition to turbulence observed in experiment $\Re=1,700\Ha$.}\label{fig2}
\end{figure}

\begin{figure}[tb]
  \includegraphics[scale=0.35]{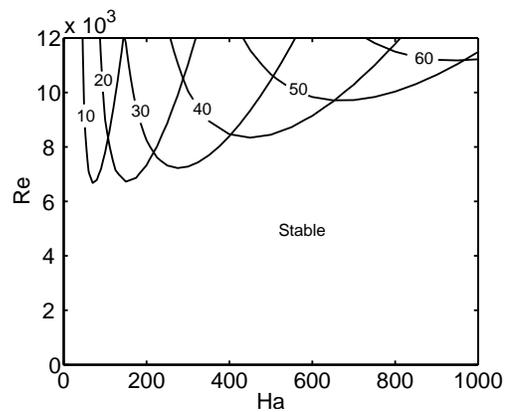}\\
  \caption{Calculated
  marginal stability curves for geometry of Obninsk experiment.
  Azimuthal mode numbers $m$ are shown. Difference between Hg and Na is within the line width.}\label{fig3}
\end{figure}

The following numerical procedure was performed for two values of
magnetic Prandtl number corresponding to mercury (Hg) and liquid
sodium (Na) and for the geometries of both Grenoble and Obninsk
experiments (Table~\ref{table1}). Taking a value of azimuthal number
from the array $m=0\div200$, we scanned through a range of values of
$\Re$ and $\Ha$, finding the maximal growth rate for given
parameters. For each value of $\Ha$ we determined the value of $\Re$
that yields a marginal stability, i. e. corresponds to the zero
maximal growth rate $\Re(\gamma)=0$. Thus we obtained the marginal
stability curves at the plane $\Re-\Ha$ (Fig.~\ref{fig2},
\ref{fig3}).

One can see from Fig.~\ref{fig2} and \ref{fig3} that the instability
threshold is determined by different azimuthal modes at different
Hartmann numbers $\Ha$. For larger $\Ha$ the corresponding $m$ is
larger. In fact, a simple scaling law can be obtained for large $m$:
$\Ha\propto m^2$. The stability curves of axisymmetric modes with
$m=0$ are not shown because they are situated at much higher values
of $\Re$. Therefore, the non-axisymmetric modes play the decisive
role for excitation of MRI in electrically driven flow.

It is worth noting here that the marginal stability curves
practically do not depend on the magnetic Prandtl number $\Pr$ of
the fluid; they are determined only by the geometry of the channel.
This is true for non-axisymmetric modes in the limit $\Pr\ll1$ which
is common for most liquid metals. In this case a good approximation
for instability threshold can be achieved by neglecting all the
terms containing $\Pr$ in the Eq. (\ref{main2}) and reducing the
system (\ref{main1}), (\ref{main2}) to one vector equation with
hydrodynamical variables. Such approach is not applicable for
axisymmetric modes in axial magnetic field \cite{PPPL4}.

In ideal MHD, the MRI threshold to a significant degree is affected
by singularities inherent in eigen-value problem \cite{Coppi2}. In
incompressible limit, the MRI of non-axisymmetric modes is
associated with one set of these singularities -- the so-called
Alfven resonances \cite{my2}. Within the frame of MHD model
considered here, the Alfven resonances are removed by dissipative
effects (resistivity and viscosity). An important consequence of
this is that the dissipative stability threshold for
non-axisymmetric modes appears to be lower than the ideal one, i. e.
the dissipation (mainly the resistivity) has destabilizing effect on
the ideal modes.

Our main observation is that the MRI threshold of the electrically
driven flow is formed by the envelope of all marginal stability
curves corresponding to modes with different azimuthal numbers $m$.
The shape of the envelope depends on the particular geometry of the
channel (see  Fig.~\ref{fig2} and \ref{fig3}). In the
Fig.~\ref{fig2}, the envelope is close to the form $\Re\propto\Ha$
when $\Ha\gtrsim300$. It means that instability should be excited in
this geometry if the ratio $\R=\Re/\Ha$ exceeds some critical value
$\R_c$. As mentioned above this effect was actually detected
experimentally \cite{Grenoble}.

For comparison of our numerical results with the experimental data
we need to calculate $\R_c$ taking into account our definition of
Reynolds number (\ref{param}). The relation between Reynolds number
$\Re^*$ from Ref. \cite{Grenoble} and $\Re$ used in our calculations
is
$$
\Re^*=\frac{dv_m}{\nu}=\frac{2LM_0}{\nu(R_2-R_1)}\ln\frac{R_2}{R_1}
\approx0.22\Re,
$$
where $d=2L$ is the channel height and $v_m$ is the mean velocity in
the equilibrium flow. Our definition of $\Ha$ is the same as
definition of $\Ha$ from Ref. \cite{Grenoble} used in the figures
(though it is not the same as stated in Eq. (2.4) in Ref.
\cite{Grenoble}). Thus we obtain
$$
\R_c=\frac{\Re}{\Ha}\approx\frac{1}{0.22}\R_c^*\approx1700.
$$
The line associated with this experimental value of $\R_c$ is also
plotted in Fig.~\ref{fig2}. As one can see, the calculated
threshold of MRI is in a good agreement with the experimental
results.

\begin{figure}[tb]
  \includegraphics[scale=0.35]{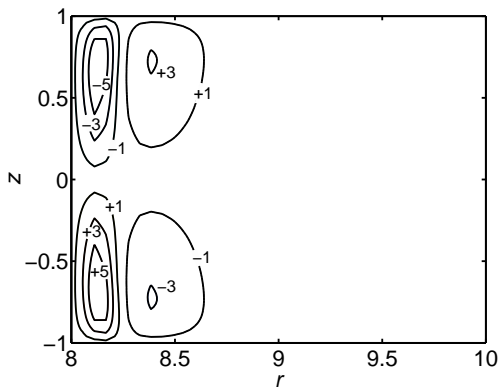}\\
  \caption{Calculated
  contours of marginally stable eigenfunction
  $v_{\varphi}$ in cross-section of Grenoble channel
  at $m=100$, $\Ha=500$, $\Re=9\cdot10^5$. Magnitude
  is shown in arbitrary units.}\label{fig4}
\end{figure}

A natural question arises: what do we observe in reality -- MRI or
Hartmann layer instability? A strong argument for MRI is the
comparison of respective linear instability thresholds: MRI
threshold found in our calculations corresponds to that measured in
the Grenoble experiment and two orders of magnitude smaller than
instability threshold of Hartmann layers \cite{Lock}. Also it is
evident that in the presence of global (affecting the entire flow)
robust linear instability such as MRI the nonlinear effects in
Hartmann layers \cite{Kras1,Kras2} are unlikely to play the major
role in destabilizing the flow. The global character of MRI is
illustrated in Fig.~\ref{fig4} where a typical marginally stable
eigenfunction $v_{\varphi}$ is shown.

In conclusion, we have shown that the instability threshold in
electrically driven flow is determined by MRI of global
non-axisymmetric modes with large azimuthal mode numbers $m$. This
threshold does not depend on the type of conducting fluid as long as
the magnetic Prandtl number of the fluid is small $\Pr\ll1$. MRI
threshold calculated for the geometry of the Grenoble experiment
agrees well with the critical ratio $\R_c$ found in this experiment.
These results suggest that the transition to turbulence  observed in
the experiment \cite{Grenoble} is associated with non-axisymmetric
MRI in electrically driven flow.

This work is supported in part by NSERC Canada.

\bibliography{MRI_expt}

\end{document}